\documentclass[twocolumn,twocolappendix]{aastex631}
\usepackage{amsmath}
\shorttitle{UV photodesorption of CH$_3$CN}
\shortauthors{Basalgete et al.}
\graphicspath{{./}{figures/}}
\begin{document}

\title{Photodesorption of acetonitrile CH$_3$CN in UV-irradiated regions of the Interstellar Medium : an experimental evidence}

\author[0000-0002-1256-7886]{Romain Basalg\`{e}te}
\affiliation{Sorbonne Universit\'{e}, Observatoire de Paris, PSL University, CNRS, LERMA, F-75014, Paris, France}

\author[0000-0002-0480-1855]{Antonio Jesus Ocaña}
\affiliation{Sorbonne Universit\'{e}, Observatoire de Paris, PSL University, CNRS, LERMA, F-75014, Paris, France}

%\collaboration{6}{(AAS Journals Data Editors)}

\author[0000-0002-8464-0536]{Géraldine F\'{e}raud}
\affiliation{Sorbonne Universit\'{e}, Observatoire de Paris, PSL University, CNRS, LERMA, F-75014, Paris, France}

\author{Claire Romanzin}
\affiliation{Univ Paris Saclay, CNRS UMR 8000, ICP, F-91405, Orsay, France}

\author[0000-0002-7405-9609]{Laurent Philippe}
\affiliation{Sorbonne Universit\'{e}, Observatoire de Paris, PSL University, CNRS, LERMA, F-75014, Paris, France}

\author[0000-0002-8441-5048]{Xavier Michaut}
\affiliation{Sorbonne Universit\'{e}, Observatoire de Paris, PSL University, CNRS, LERMA, F-75014, Paris, France}

\author[0000-0003-3200-4113]{Jean-Hugues Fillion}
\affiliation{Sorbonne Universit\'{e}, Observatoire de Paris, PSL University, CNRS, LERMA, F-75014, Paris, France}

\author[0000-0002-9021-2415]{Mathieu Bertin}
\affiliation{Sorbonne Universit\'{e}, Observatoire de Paris, PSL University, CNRS, LERMA, F-75014, Paris, France}



\begin{abstract}
   Pure acetonitrile (CH$_3$CN) and mixed CO:CH$_3$CN and H$_2$O:CH$_3$CN ices have been irradiated at 15 K with Vacuum UltraViolet (VUV) photons in the 7-13.6 eV range using synchrotron radiation. VUV photodesorption yields of CH$_3$CN and of photo-products have been derived as a function of the incident photon energy. The coadsorption of CH$_3$CN with CO and H$_2$O molecules, which are expected to be among the main constituents of interstellar ices, is found to have no significant influence on the VUV photodesorption spectra of CH$_3$CN, CHCN$\bullet$, HCN, CN$\bullet$ and CH$_3\bullet$. Contrary to what has generally been evidenced for most of the condensed molecules, these findings point toward a desorption process for which the CH$_3$CN molecule that absorbs the VUV photon is the one desorbing. It can be ejected in the gas phase as intact CH$_3$CN or in the form of its photo-dissociation fragments. Astrophysical VUV photodesorption yields, applicable to different locations, are derived and can be incorporated into astrochemical modeling. They vary from $0.67 (\pm 0.33) \times 10^{-5}$ to $2.0 (\pm 1.0) \times 10^{-5}$ molecule/photon for CH$_3$CN depending on the region considered, which is high compared to other organic molecules such as methanol. These results could explain the multiple detections of gas phase CH$_3$CN in different regions of the interstellar medium and are well-correlated to astrophysical observations of the Horsehead nebula and of protoplanetary disks (such as TW Hya and HD 163296).
\end{abstract}



\section{Introduction} \label{sec:intro}

Acetonitrile (CH$_3$CN) is a complex organic molecule (COM) regularly detected in several regions of the Interstellar Medium (ISM) at different stages of star formation \citep{purcell_ch3cn_2006, bergner_survey_2018, calcutt_alma-pils_2018, Liszt_2018, zeng_complex_2018}. In the ISM, acetonitrile can be formed both by gas phase reactions and by grain surface chemistry in icy mantles. In protoplanetary disks, pure gas phase chemistry fails to explain the observed abundances of CH$_3$CN \citep{oberg_comet-like_2015, loomis_distribution_2018}, indicating that its detection is intimately linked to processes associated with ices. In these disks, CH$_3$CN detection extends far from the central Young Stellar Object (YSO), in regions where the temperature of the icy mantles is expected to be low enough ($T < 100$ K) so that gas phase CH$_3$CN should accrete on the surface of the icy dust grains. The presence of CH$_3$CN in the gas phase therefore implies that a non-thermal process should maintain a sufficient amount of gas phase CH$_3$CN via desorption from the surface of ice mantles. UV photodesorption of CH$_3$CN from ices, which is the desorption induced by the UV irradiation of ices, is so far the most important non-thermal process taken into account in disk modeling. In the case of the Horsehead nebula, for which the separation between the UV-illuminated Photo-Dominated Region (PDR) and the UV-shielded dense core is easily made \citep{gratier_iram30_2013, guzman_2014}, observations show that CH$_3$CN gas abundance is 30 times higher in the PDR region than in the dense core. As pure gas phase reactions also fail to explain these abundances, UV photodesorption of CH$_3$CN from icy mantles is expected to play a major role in PDRs.
\\\\
Recent experimental investigations on CH$_3$CN Vacuum UV photodesorption were conducted using an indirect method in the 7-10.2 eV range \citep{bulak_novel_2020}. UV photodesorption of CH$_3$CN from pure CH$_3$CN ice was estimated to be $< 7.4 \times 10^{-4}$ molecule desorbed by incident photon. This study is so far the only constraint on the UV photodesorption of CH$_3$CN from pure CH$_3$CN ices. Several experimental studies regarding UV photodesorption of simple molecules such as CO, CO$_2$, N$_2$ or CH$_4$ \citep{bertin_indirect_2013,fillion_wavelength_2014,carrascosa_13co_2019, dupuy_spectrally-resolved_2017} and more complex molecules such as CH$_3$OH and H$_2$CO \citep{bertin_uv_2016, martin-domenech_study_2016, feraud_vacuum_2019} have shown that UV photodesorption yields are strongly dependent on the ice composition: e.g., when diluted in CO-dominated ices, CH$_3$OH UV photodesorption is quenched whereas H$_2$CO UV photodesorption is enhanced, compared to the case of pure CH$_3$OH and H$_2$CO ice respectively. This indicates that the UV photodesorption of complex molecules should be systematically studied as a function of the ice composition. This is even more relevant when considering that pure ices of organics are not likely to be found in the ISM. However, the exact role played by the ice composition is still difficult to predict a priori. We thus propose to provide absolute photodesorption yields in composite ices where the molecule to study is embedded in a CO or H$_2$O icy matrix. Those two kinds of ices can serve as models of the realistic icy mantles that could be found in the ISM, mainly composed of water, but which can also present a CO-rich phase in denser regions beyond the CO snowlines \citep{pontoppidan_2003, boogert_observations_2015}.
\\\\
In the present study, we quantify the Vacuum UV (VUV) photodesorption, in the 7-13.6 eV range, of CH$_3$CN from pure CH$_3$CN ice and from CH$_3$CN mixed with or on top of CO and H$_2$O ices. VUV photodesorption yields of photo-products are also derived. The yields are obtained as a function of the incident VUV photon energy, which allows to discuss the physical mechanisms at play. Absolute VUV photodesorption yields applicable to astronomically relevant environments are also provided in order to be incorporated into astrochemical modeling. In Sect.\ref{sec:methods}, the experimental procedure is presented. In Sect.\ref{results}, the results and their astrophysical implications are discussed.

\section{Methods} \label{sec:methods}

Experiments are conducted at the DESIRS beamline of the SOLEIL synchrotron facility \citep{Nahon:ie5079}. The setting used on this beamline provides VUV photons in the energy range of 7 - 14 eV with a $\sim 1$ eV bandwidth. The photon flux, measured with a calibrated silicon photodiode, depends on the photon energy. It varies from $6.3 \times 10^{14}$ photon.s$^{-1}$ at 7 eV to $6.4 \times 10^{13}$ photon.s$^{-1}$ at 14 eV. The beamline is equipped with a rare gas filter to suppress higher energy harmonics from the undulator. The SPICES setup (Surface Processes and ICES) is connected to the synchrotron beamline to run the experiments. It consists of an ultra-high vacuum chamber (base pressure $\sim 10^{-10}$ mbar) within which a rotatable copper substrate (polycrystalline oxygen-free high-conductivity copper) can be cooled down to T$\sim$15 K by a closed-cycle helium cryostat. A tube positioned a few millimeters away from the substrate allows to inject a partial pressure of gas phase molecules that will stick on the cold substrate, thus forming the molecular ices. During this process, the base pressure of the chamber is not significantly modified. The gaseous molecules used during these experiments were acetonitrile CH$_3$CN (99.8\% purity, Sigma-Aldrich), carbon monoxide CO (Air Liquide, 99.9\%) and water H$_2$O (liquid chromatography standard, Fluka). Liquid products (H$_2$O and CH$_3$CN) were further purified using several freeze-pump-thaw cycles. To form the mixed ices (CO:CH$_3$CN and H$_2$O:CH$_3$CN), the corresponding gaseous mixture is prepared in a gas-introduction system equipped with a capacitive pressure gauge before introduction into the experiment. 
\\\\
The ice thickness is expressed in monolayer (ML), roughly equivalent to a molecule surface density of $10^{15}$ molecules.cm$^{-2}$. Prior to the synchrotron experiments, the temperature-programmed desorption (TPD) technique is used to calibrate the ice thickness. This calibration allows us to control the number of ML deposited onto the substrate with a relative precision of about 10 \% \citep{bertin_nitrile_2017}.  Pure CH$_3$CN ices ($\sim 60$ ML) and mixed CO:CH$_3$CN ($\sim 100$ ML) ices were formed at 15 K. Mixed H$_2$O:CH$_3$CN ices ($\sim 100$ ML) were formed at 100 K and cooled down to 15 K before irradiation to ensure that the resulting water ice is in its compact amorphous phase. This phase is commonly referred to as compact amorphous solid water (c-ASW). 1 ML of CH$_3$CN was also deposited on pure CO ice ($\sim$ 30 ML) at 15 K and on pure H$_2$O ice ($\sim$ 30 ML) at 100 K, followed by a cooling down to 15 K. 
\\\\
The VUV beam is sent at a 45$^{\circ}$ incidence on the ice surface on a spot of $\sim$ 1 cm$^{2}$. While the ices are irradiated, the photodesorption of neutral species is monitored by recording the desorbed molecules in the gas phase using a quadripolar mass spectrometer (QMS) equipped with an electron-impact ionization chamber at 70 eV. Each gas-phase species is probed by monitoring the mass signal of its corresponding intact cation. Typical desorption signals (raw data) are shown in the Appendix \ref{AppA}. At the end of the irradiation procedure, TPD experiments (from 15 K to 200 K with a ramp of 12 K.min$^{-1}$) are performed to evaporate all molecules from the substrate before a new ice is formed. These TPD experiments have also been used to investigate any photochemistry (see Appendix \ref{AppB}). The photodesorption yields presented in this article (in molecule desorbed per incident photon, displayed as molecule/photon for more simplicity in the following) are derived from the raw desorption signals given by the QMS on the considered mass channel. A detailed explanation is provided in the Appendix \ref{AppA}. The mixture ratios used for the mixed ices were chosen to progressively study the impact of the dilution of CH$_3$CN in CO or H$_2$O ices on the photodesorption yields but may not be representative of the interstellar ice analogs that could be found in the ISM.

\begin{figure*}[!ht]
\begin{center}
\includegraphics[width=17cm]{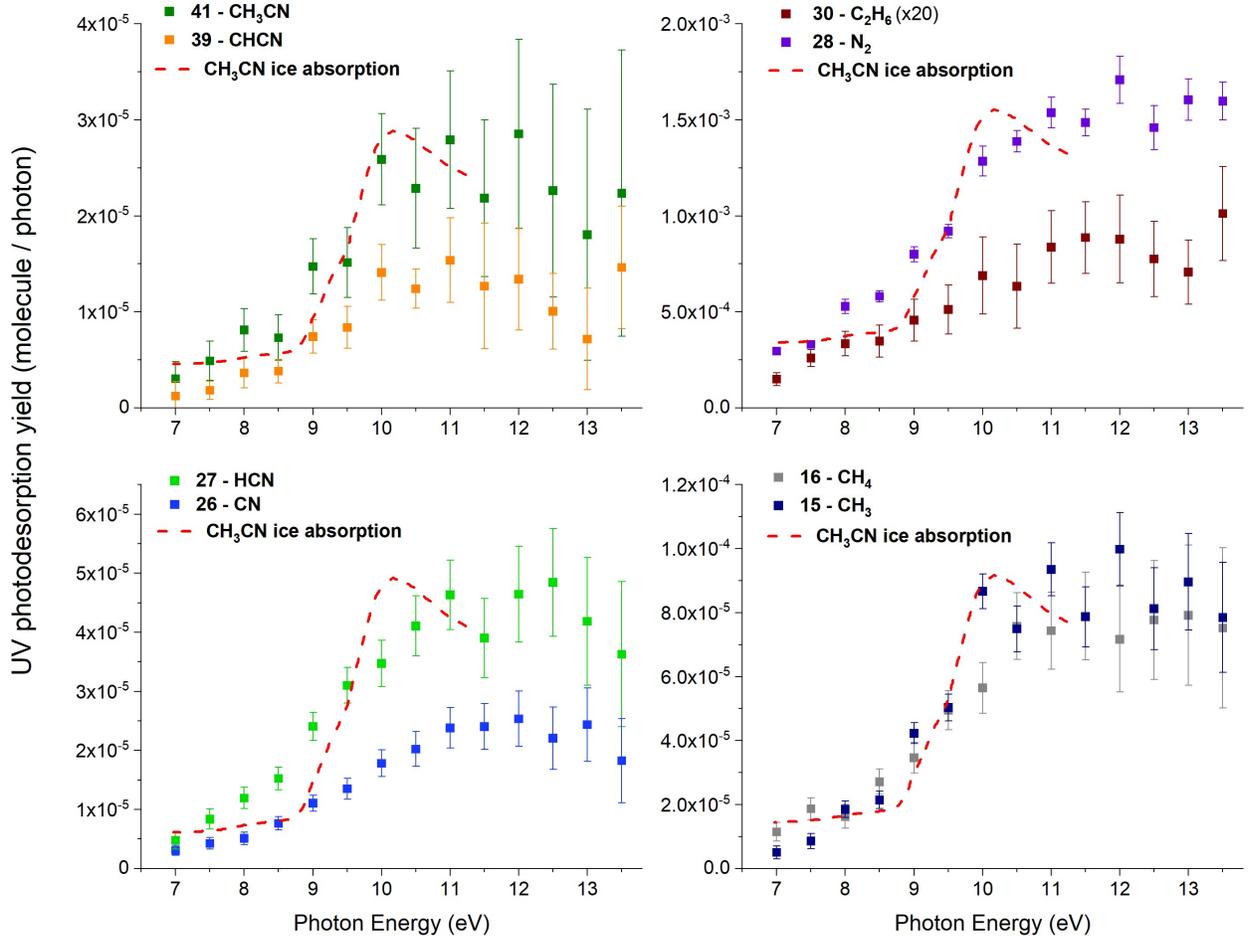}
\caption{VUV photodesorption spectra (squares with error bars), as a function of the incident photon energy, of CH$_3$CN (mass 41), CHCN (mass 39), C$_2$H$_6$ (mass 30), N$_2$ (mass 28), HCN (mass 27) , CN (mass 26), CH$_4$ (mass 16) and CH$_3$ (mass 15) from pure CH$_3$CN ice at 15 K. Each mass channel is corrected from the cracking of higher masses fragmentation in the mass spectrometer. The attribution of species to the mass channels is discussed in Sect.\ref{sec:pure}. In each panel, we also displayed in red dashed lines the VUV absorption spectrum of CH$_3$CN ice at 10 K (in arbitrary units) from \citealt{sivaraman_vacuum_2016}.}
\label{Graph_Pur}
\end{center}
\end{figure*}

\begin{figure*}[!ht]
\begin{center}
\includegraphics[width=17cm]{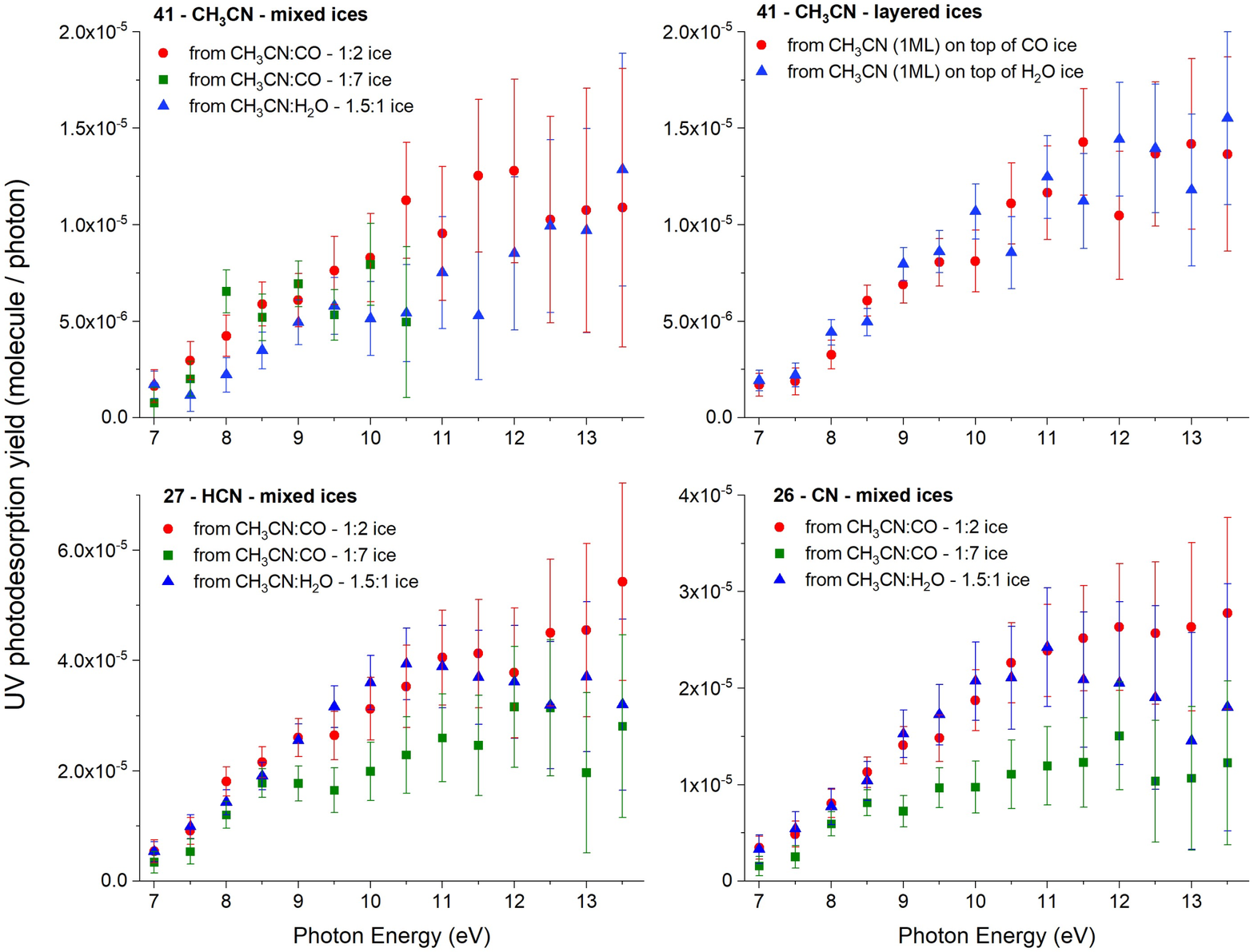}
\caption{VUV photodesorption spectra, as a function of the incident photon energy, of CH$_3$CN (mass 41) from mixed CH$_3$CN:CO and mixed CH$_3$CN:H$_2$O ices (top left pannel) and from 1 ML of CH$_3$CN on top of CO and on top of H$_2$O ice (top right pannel). VUV photodesorption of HCN (mass 27) and CN (mass 26) from mixed CH$_3$CN:CO and CH$_3$CN:H$_2$O ices are presented in the bottom panels. These yields were derived for ices at 15 K.}
\label{Graph_mix}
\end{center}
\end{figure*}

\section{Results and discussion} \label{results}
\subsection{Pure acetonitrile ice} \label{sec:pure}
In Figure \ref{Graph_Pur}, we present the VUV photodesorption spectra derived from our experiments on pure CH$_3$CN ices and we also compare these spectra with the VUV absorption spectrum (in the 7 - 11.2 eV range) of pure CH$_3$CN ice at 10 K from \citealt{sivaraman_vacuum_2016}. The attribution of photo-products to the mass channels is discussed in the following. The VUV photodesorption of masses 15 and 26 is attributed to CH$_3\bullet$ and CN$\bullet$ respectively, which are expected to be direct UV photo-dissociation product of CH$_3$CN. This is consistent with the UV photo-dissociation of gas phase acetonitrile for which C-C bond breaking is observed \citep{moriyama_vacuum_1998, kanda_photodissociation_1999}. HCN and CH$_4$ are observed as major photo-products in VUV photolysis experiments of pure CH$_3$CN ice at 20 K \citep{bulak_photolysis_2021} and 12 K \citep{hudson_reactions_2004}. Accordingly, we attributed the masses 16 and 27 to CH$_4$ and HCN photodesorption respectively. HCN was also proposed as a possible direct UV photo-dissociation product of CH$_3$CN in gas phase by \citealt{schwell_vuv_2008}. We assumed that the mass 39 corresponds to CHCN$\bullet$ photodesorption. Based on gas phase experiments for which H-loss reaction is assumed to occur for the UV photo-dissociation of CH$_3$CN \citep{moriyama_vacuum_1998, schwell_vuv_2008}, the detection of CHCN$\bullet$ in our experiment could be expected to come from dehydrogenation of CH$_3$CN induced by VUV photo-absorption.  A raw photodesorption signal was detected on the mass 40, which could correspond to the desorption of CH$_2$CN$\bullet$. However, when considering the fragmentation of the mass 41 (CH$_3$CN) into the mass 40 (CH$_2$CN$\bullet$) in the ionization chamber of our QMS, the corrected signal on the mass 40 falls below our detection limit so that we can only provide an upper limit of $\sim 1 \times 10^{-6}$ molecule/photon for the VUV photodesorption yield of CH$_2$CN$\bullet$ from pure acetonitrile ices (this is also the case for the binary mixed ices in Section \ref{sec:mix}).
\\\\
Additionally, \citealt{hudson_reactions_2004} has observed the isomerization of CH$_3$CN into CH$_3$NC during VUV irradiation (with an hydrogen lamp) of pure CH$_3$CN ice at 12 K. This isomerization was detected by infrared via the -N$\equiv$C streching mode at 2170 cm$^{-1}$. From \citealt{hudson_reactions_2004}, we could expect to have only a few percent of CH$_3$CN isomerization during our irradiation experiments and we did not observe the -N$\equiv$C streching mode by infrared on pure CH$_3$CN ices (this is also the case for the binary mixed ices in Section \ref{sec:mix}). Therefore, we made the assumption that the VUV photodesorption of the mass 41 in our experiments is dominated by CH$_3$CN desorption rather than CH$_3$NC. The isomerization of HCN into HNC was not observed by \citealt{hudson_reactions_2004} and accordingly, we estimated that the VUV photodesorption of the mass 27 in our experiments is dominated by HCN desorption and not its isomer HNC.
\\\\
The mass 30 was attributed to C$_2$H$_6$ desorption, assumed to originate from radical-radical reaction between CH$_3\bullet$ molecules, based on low-energy electron irradiation experiments of pure CH$_3$CN ices \citep{ipolyi_2007, bass_reactions_2012}. The mass 28 is the most desorbing one in our experiments. It could be attributed to N$_2$, H$_2$CN and/or C$_2$H$_4$ VUV photodesorption. Post-irradiation TPD experiments have also revealed a mass 28 desorption features which peaks below 40 K, and intensity of which increases with the total irradiation fluence; and not observed on non-irradiated ices (see Figure \ref{Graph_tpd} in the Appendix \ref{AppB}). This indicates that the irradiation also implies the accumulation of the associated species onto the ice during irradiation. Among all the possible candidates at the origin of the mass 28 channel, the low desorption temperatures points toward its association with N$_2$ (sub)monolayer desorption \citep{collings_laboratory_2004, smith_desorption_2016}. Accordingly, we assumed that the photodesorption of the mass 28 from pure CH$_3$CN ice was dominated by N$_2$. The formation of N$_2$ might result from photochemistry involving CN$\bullet^{*}$ which is left in electronically excited states (A$^{2} \Pi$, B$^{2} \Sigma ^{+}$) after dissociation of CH$_3$CN according to gas phase experiments \citep{schwell_vuv_2008}.
\\\\
The VUV photodesorption spectra of the species presented in Figure \ref{Graph_Pur} have the same dependence in energy: the photodesorption efficiency is increasing from 7 to 10-10.5 eV and seems to be constant above 10.5 eV. The small variations above this energy are significantly lower than the error bars, and cannot be confidently considered as a "real" modification of the photodesorption yields. In Figure \ref{Graph_Pur}, we can also see that there is a good correlation between the VUV absorption spectrum of CH$_3$CN ice and the VUV photodesorption spectra of the different molecules presented. The similarities between these spectra indicates that the photodesorption of these species originates from the same phenomena, which is the VUV absorption of CH$_3$CN in condensed phase. Concerning the photodesorption efficiencies, they range from $\sim 1.2 \times 10^{-5}$ molecule/photon for CHCN$\bullet$ to $\sim 1.6 \times 10^{-3}$ molecule/photon for N$_2$, around 10.5 eV. Finally, the VUV photodesorption yield of intact CH$_3$CN is $\sim 2.5 \times 10^{ -5}$ molecule/photon at 10.5 eV. Our results compare very well to the recent study of \citealt{bulak_novel_2020} who could not measure directly the photodesorption yield of CH$_3$CN from pure CH$_3$CN ice but provided an upper limit of $7.4 \times 10^{-4}$ molecule/photon, compatible with our measurements.

\subsection{Binary ices}\label{sec:mix}
\begin{figure}[!ht]
\begin{center}
\resizebox{\hsize}{!}{\includegraphics[width=17cm]{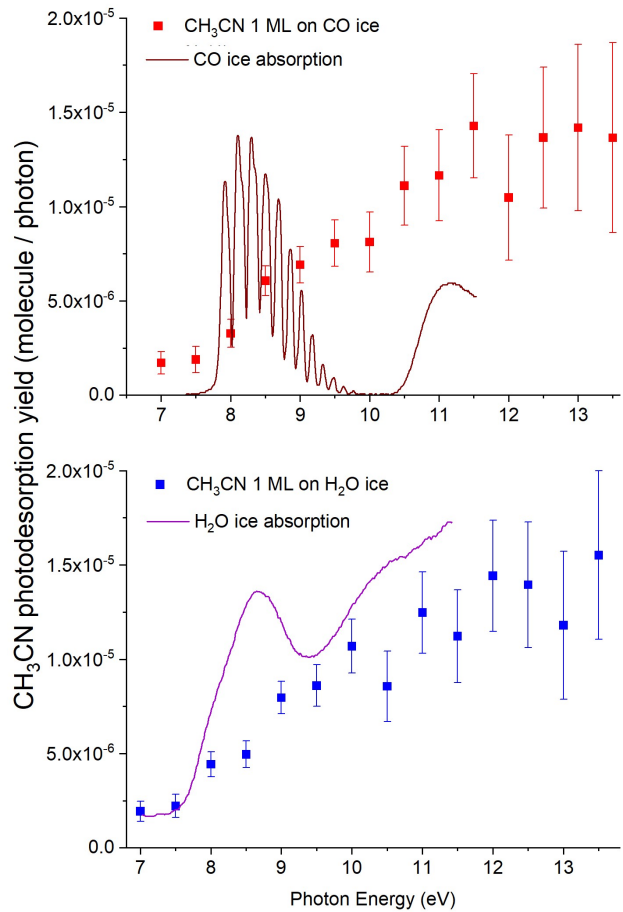}}
\caption{Top panel: comparison of CH$_3$CN VUV photodesorption spectrum from a layered CH$_3$CN (1 ML) on top of CO ice with the VUV absorption spectrum of CO in condensed phase (from \citealt{Lu_2005}). Bottom pannel: comparison of CH$_3$CN VUV photodesorption spectrum from a layered CH$_3$CN (1 ML) on top of H$_2$O ice with the VUV absorption spectrum of H$_2$O in condensed phase (from \citealt{lu_absorption_2008}). The VUV absorption spectrum of CO and H$_2$O in condensed phase are shown in arbitrary units.}
\label{Graph_ind}
\end{center}
\end{figure}

In Figure \ref{Graph_mix}, we summed up our results on CH$_3$CN, HCN and CN$\bullet$ photodesorption from the binary ices tested : mixed CH$_3$CN:CO ice, mixed CH$_3$CN:H$_2$O ice and layered ices ($\sim 1$ ML of CH$_3$CN on top of pure CO and on top of pure H$_2$O ice). For the specific case of CH$_3$CN photodesorption from a mixed CH$_3$CN:CO (1:7) ice, the yields above 10.5 eV are not shown for more visibility because of the large error bars. In addition to the presented molecules in Figure \ref{Graph_mix}, we searched for the desorption of masses 42, 43, 45 and 52 which could correspond to CNO$\bullet$, HCNO, NH$_2$CHO (formamide) and C$_2$N$_2$ respectively. However, the signals on these channels were too noisy to derive a photodesorption yield and we can only provide an upper limit of $5 \times 10^{-6}$ molecule/photon for these masses. The photodesorption data for the masses 30, 28 and 16 in the case of the mixed ices are displayed in the Appendix \ref{AppC} in Figure \ref{m_302816}. The desorption signals for the masses 30 and 28 are significantly higher in the case of mixed CH$_3$CN:CO ices compared to the case of mixed CH$_3$CN:H$_2$O ice. This is due to a contribution of CO photodesorption for the mass signal 28 and most probably H$_2$CO and/or NO photodesorption for the mass signal 30 in the case of mixed CH$_3$CN:CO ices. The desorption signal on the mass 16 is not significantly dependent on the presence of CO or H$_2$O molecules in the ice but a possible contribution of atomic O photodesorption to this mass signal prevents us from precisely quantifying a possible VUV photodesorption of CH$_4$ from the mixed ices.
\\\\
We do not observe a significant influence of the presence of H$_2$O and CO molecules on the shape of the VUV photodesorption spectra for the presented molecules in Figure \ref{Graph_mix} (CH$_3$CN, HCN, CN$\bullet$). A similar behaviour is observed for the masses 15 and 39 which can be unambiguously attributed to CH$_3\bullet$ and CHCN$\bullet$ (see Figure \ref{m_39_15} in the Appendix \ref{AppC}). In similar experiments, indirect desorption mechanisms have been observed in the case of simpler molecules such as N$_2$ or CH$_4$ deposited on top of pure CO ice \citep{bertin_indirect_2013, dupuy_spectrally-resolved_2017}, "pre-irradiated" pure CO$_2$ ice \citep{fillion_wavelength_2014} or N$_2$, CO, Kr and Ar on top of pure water c-ASW ice \citep{PhysRevLett.126.156001}. In these cases, the desorption of one molecule is induced by the UV photo-absorption of another one whose excited state relaxation induces an energy transfer to the desorbing molecule. Experimentally, this results in similarities between the VUV photodesorption spectrum of the molecule considered and the VUV absorption spectrum of CO or H$_2$O in condensed phase. In the case of more complex molecules, such as CH$_3$OH, this indirect mechanism was not observed \citep{bertin_uv_2016}. In our CH$_3$CN experiments, the VUV photodesorption spectra of CH$_3$CN, CHCN$\bullet$, HCN, CH$_3\bullet$ and CN$\bullet$ from the mixed and layered ices do not show any similarities with the UV absorption profile of CO or H$_2$O molecules in condensed phase. These data are compared for the case of CH$_3$CN UV photodesorption from the layered ices in Figure \ref{Graph_ind}. It is however surprising to observe that the VUV photodesorption yields of CH$_3$CN from CO-dominated ices are similar (when considering the error bars) between the 1:2 and 1:7 ratios (see Figure \ref{Graph_mix}, upper left panel). As CH$_3$CN is more diluted in the CH$_3$CN:CO - 1:7 ice, we would have expected to obtain a lower yield compared to the case of the CH$_3$CN:CO - 1:2 ice. The fact that there is no clear signature of the CO UV absorption spectrum between 7.5 and 9.5 eV in the photodesorption spectrum of CH$_3$CN from CH$_3$CN (1 ML) on top of CO ice (see Figure \ref{Graph_ind}) seems to exclude the possibility of a significant indirect desorption mechanism induced by CO molecules when the CH$_3$CN surface concentration is high. However, for higher dilution of CH$_3$CN in a CO environment, a desorption of CH$_3$CN triggered by the excitation of the CO matrix cannot be totally ruled out. The similarities of the photodesorption yields of CH$_3$CN from CO-dominated ices for the 1:2 and the 1:7 ratios could also be explained by less bounded CH$_3$CN molecules in CO-dominated ices but this point would need additional experiments in lower concentration regimes to conclude. Finally, an indirect desorption mechanism induced by CO or H$_2$O molecules does not seem to be at play or significant in our set of experiments for the desorption of the molecules cited above (CH$_3$CN, CHCN$\bullet$, HCN, CH$_3\bullet$ and CN$\bullet$) from CH$_3$CN-containing ices.\\

The absolute values of the VUV photodesorption yields of CH$_3$CN, CHCN$\bullet$, HCN, CH$_3\bullet$ and CN$\bullet$ from the binary mixed ices are slightly diminished compared to the case of pure CH$_3$CN ice but this effect is not significant when considering the error bars and could be explained by a lower number of CH$_3$CN molecules available for desorption at the ice surface, inducing a lower desorption flux in the case of mixed ices. The deposition technique (mixed ices or layered ices) does not have a significative influence on the VUV photodesorption of these molecules either. The photodesorption yields of CH$_3$CN from 1 ML of CH$_3$CN on top of CO or H$_2$O ice are about half of that from pure CH$_3$CN ice, which could be due to the fact that several CH$_3$CN-containing layers are involved in the desorption of CH$_3$CN from pure CH$_3$CN ice. It is much likely that the VUV photodesorption of CH$_3$CN, CHCN$\bullet$, HCN, CH$_3\bullet$ and CN$\bullet$ from the ices tested (pure, mixed and layered ices) occurs via a process for which the CH$_3$CN molecule that absorbs the VUV photon is the one desorbing. It can be ejected as intact CH$_3$CN or in the form of its photo-dissociation fragments. The desorption of CH$_3$CN may involve dissociation and exothermic recombination of the excited molecule. According to our results, this process seems to be independent of the presence of CO or H$_2$O molecules surrounding the VUV absorbing CH$_3$CN molecule in the ice. 

\subsection{Astrophysical implications}
By integrating our experimental photodesorption spectra with UV spectra that are representative of the UV fields found in different regions of the ISM, we are able to provide photodesorption yields that are suitable for astrophysical environments. The UV field used were: the UV emission spectra of a classical T Tauri Star representative of protoplanetary disks \citep{france_high-resolution_2014}, the UV field produced by cosmic-ray secondary electrons representative of dense interstellar clouds \citep{Gredel_1987} and the UV interstellar radiation field (ISRF) from \citet{Mathis_1983}. The resulting yields are presented in Table \ref{Astro_Yields} and can be incorporated in astrochemical modeling. As discussed previously, we found that the VUV photodesorption of CH$_3$CN and what is associated with its photo-fragments (CHCN$\bullet$, HCN, CN$\bullet$ and CH$_3\bullet$) does not strongly depend on the presence of CO or H$_2$O molecules in the ice, which are expected to be among the main constituents of ices in the ISM \citep{boogert_observations_2015}. Accordingly, we took the photodesorption spectra from pure CH$_3$CN ice (Figure \ref{Graph_Pur}) to derive the yields presented in Table \ref{Astro_Yields}, which are thus assumed to be applicable to any CO or H$_2$O dominated ices containing CH$_3$CN molecules. The yields corresponding to protoplanetary disks and dense interstellar clouds are found to be identical within the error bars and are approximatively equal to the experimental yields at the Lyman $\alpha$ around 10.2 eV (Figure \ref{Graph_Pur}), as it is the energy that dominates the UV field in these regions. The yields corresponding to the ISRF are found to be lower due to a higher contribution of the UV field at low energy ($<$ 10 eV) where photodesorption is less efficient. Interestingly, the desorption of CH$_3$CN and its photo-fragments are found in the same order of magnitude in the region considered. The desorption of the masses 30, 28 and 16 from the mixed ices will not be discussed in this section as it is not possible to clearly attribute species to these mass channels (see Sect.\ref{sec:mix}). Moreover, the formation of the associated species (N$_2$, C$_2$H$_6$, CH$_4$) has much likely involved the absorption of at least 2 VUV photons, in a limited time and spatial interval for the resulting radicals to react. This makes their possible experimental desorption not astrophysically relevant when considering the very low UV flux observed in the discussed regions of the ISM compared to our experimental UV flux.
\\
\begin{deluxetable}{ccc}
\tablecaption{Astrophysical photodesorption yield ($\times 10^{-5}$ molecule/photon) extrapolated from our experimental results on pure CH$_3$CN ices at 15 K, for different astrophysical environments\label{Astro_Yields}}
\tablewidth{0pt}
\tablehead{
\colhead{Photodesorbed} & \colhead{ISRF$^{(a)}$} &  \colhead{Dense Clouds$^{(b)}$} \\
\colhead{species} &  &  \colhead{and Disks$^{(c)}$} 
}
\startdata
CH$_3$CN & $0.67 \pm 0.33$ & $2.0 \pm 1.0$ \\
CHCN & $0.34 \pm 0.17$ & $1.0 \pm 0.5$ \\
HCN & $1.1 \pm 0.6$ & $3.3 \pm 1.7$ \\
CN & $0.60 \pm 0.30$ & $1.7 \pm 0.9$ \\
CH$_3$ & $2.1 \pm 1.0$ & $6.5 \pm 3.3$ \\
\enddata
\tablecomments{(a) UV field from \citet{Mathis_1983}. (b) UV field from \citet{Gredel_1987}. (c) UV field from \citet{france_high-resolution_2014}}
\end{deluxetable}
\\
As stated in the Introduction, the astrophysical observations of gas phase acetonitrile CH$_3$CN and of gas phase methanol CH$_3$OH in the Horsehead UV-illuminated PDR and UV-shielded dense core region show that UV photodesorption of these molecules from icy grains should play a significant role \citep{gratier_iram30_2013, guzman_2014}. A quantitative comparison of CH$_3$OH and CH$_3$CN abundances in this region is to be put in contrast with our experimental results. According to \citealt{gratier_iram30_2013} and \citealt{guzman_2014}, CH$_3$OH abundances are found to be similar between the PDR and the dense core region whereas CH$_3$CN is 30 times more abundant in the PDR than in the dense core region. In the PDR, CH$_3$OH is found to be less abundant than CH$_3$CN by approximatively one order of magnitude. Among the possible mechanisms that could explain these observations, our experiments clearly point toward a higher quantitative role of CH$_3$CN VUV photodesorption compared to the case of CH$_3$OH VUV photodesorption. In fact, we found that VUV photodesorption of CH$_3$CN is not significantly dependent on the ice composition and is above $10^{-5}$ molecule/photon whereas, in similar experiments \citep{bertin_uv_2016}, CH$_3$OH VUV photodesorption is not detected when methanol is mixed with CO ice (only an upper limit of $10^{-6}$ molecule/photon has been derived). The higher VUV absorption cross section of CH$_3$CN compared to CH$_3$OH (based on gas phase absorption cross sections; \citealt{cheng_experimental_2002}, \citealt{schwell_vuv_2008}) along with the resilience of nitriles to UV photolysis, which has been demonstrated in the case of acetonitrile versus acetic acid in pure ices at 15 K \citep{bernstein_lifetimes_2004}, could be an interesting route to explain these experimental results. 
\\\\
Our experimental results may also explain the astrophysical observations of gas phase CH$_3$CN and CH$_3$OH in protoplanetary disks. In the case of the protoplanetary disk around the T Tauri star TW Hya, gas phase abundances of CH$_3$CN and CH$_3$OH are found to be similar \citep{walsh_first_2016, loomis_distribution_2018} whereas in the case of the disk around the Herbig Ae star HD 163296, CH$_3$CN is strongly detected and CH$_3$OH emission lines are below the detection limits \citep{bergner_survey_2018, carney_upper_2019}. As suggested in \citealt{bergner_survey_2018}, Herbig Ae stars are expected to be stronger UV emitters than T Tauri stars. Therefore, the role of VUV photodesorption of CH$_3$CN and CH$_3$OH from interstellar ices has been raised by \citealt{bergner_survey_2018} as a possible candidate to explain these differences in the observations of these COMs in protoplanetary disks. Our experimental results, that show a higher VUV photodesorption efficiency of CH$_3$CN from interstellar ice analogs, by at least one order of magnitude, compared to that of CH$_3$OH, clearly support the assumption of \citealt{bergner_survey_2018}. However, the VUV photodesorption yields of COMs used in disk modeling are often higher than the experimental ones (e.g., \citealt{loomis_distribution_2018} considered a CH$_3$CN VUV photodesorption yield of $10^{-3}$ molecule/photon in their modeling of the TW Hya disk) and other processes may be of importance to explain the detection of gaseous COMs in these regions \citep{dupuy_x-ray_2018,dartois_non-thermal_2019,ciaravella_x-ray_2020,basalgete_complex_2021-1,basalgete_complex_2021}. 
\\\\
In this experimental study, we derived a CH$_3$CN VUV photodesorption yield of $\sim 2.5 \times 10^{-5}$ molecule/photon at 10.5 eV. This photodesorption was observed to be independent of the presence of CO and H$_2$O molecules in the ice, which are expected to be among the main constituents of interstellar ices. Interestingly, CH$_3$CN VUV photodesorption is found to be more efficient than CH$_3$OH VUV photodesorption \citep{bertin_uv_2016} by at least one order of magnitude. This shows that the VUV photodesorption of organics from interstellar ices cannot be so far extrapolated from the specific case of one organic molecule. The results obtained from our experiments could participate into explaining the multiple detections of CH$_3$CN in UV-irradiated regions of the ISM, especially in the case of protoplanetary disks \citep{loomis_distribution_2018, bergner_survey_2018} and in the case of the Horsehead Nebulae \citep{gratier_iram30_2013, guzman_2014}. 
\begin{acknowledgments}
This work was done with financial support from (i) the Region Ile-de-France DIM-ACAV+ program, (ii) the Sorbonne Université "Emergence" program, (iii) the ANR PIXyES project, grant ANR-20-CE30-0018 of the French "Agence Nationale de la Recherche" and (iv) the Programme National “Physique et Chimie du Milieu Interstellaire” (PCMI) of CNRS/INSU with INC/INP co-funded by CEA and CNES. We would like to acknowledge SOLEIL for provision of synchrotron radiation facilities under Project Nos. 20191298, and we gratefully thank L. Nahon for his help on the DESIRS beamline.
\end{acknowledgments}

\bibliographystyle{aasjournal}
\bibliography{Bibli}
\appendix
\section{Derivation of the VUV photodesorption yields and determination of the error bars} \label{AppA}
\begin{figure*}
\begin{center}
\includegraphics[width=17cm]{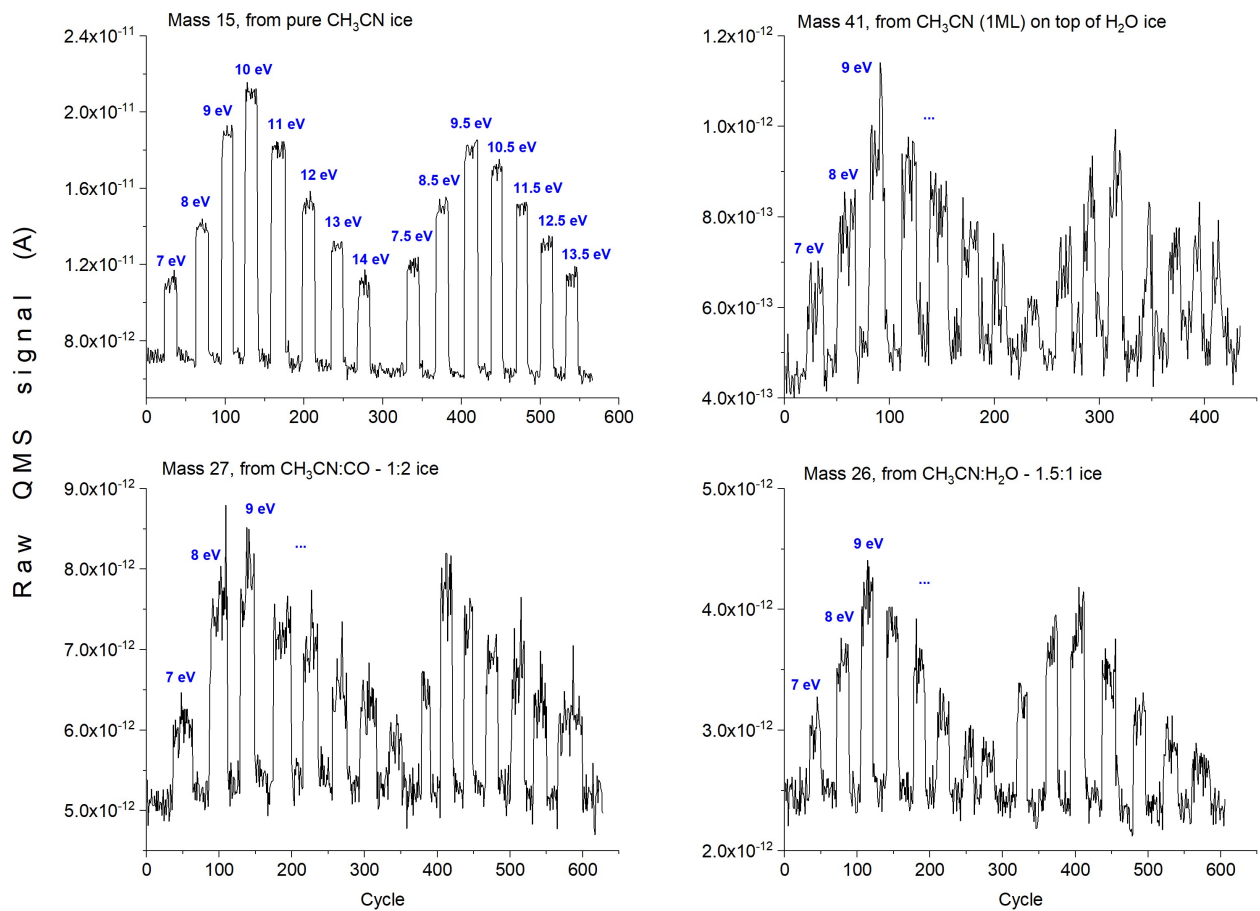}
\caption{Examples of raw QMS signal observed during irradiation experiments for different mass channels and different ices tested at 15 K. In blue is displayed the incident photon energy used for each irradiation step.}
\label{Graph_raw}
\end{center}
\end{figure*}

The photodesorption yields presented in this article (in molecule desorbed per incident photon, displayed as molecule/photon for more simplicity) are derived from the raw desorption signals given by the QMS on the considered mass channel, using the following formula :
\begin{equation}
\Gamma_X(E) = k_X \times \frac{I_X(E)}{\phi(E)}
\label{Gamma_A}
\end{equation}
where $X$ is the neutral species associated with the mass channel considered, $E$ is the photon energy, $\Gamma_X(E)$ is the photodesorption yield of $X$ at $E$, $I_X(E)$ is the current given by the QMS for the considered mass channel during the irradiation at $E$, $\phi(E)$ is the photon flux at $E$ and $k_X$ is a proportionality factor between the molecular desorption flux and the QMS signal. The ices studied in this article were sequentially irradiated during a few seconds at several fixed energies while the desorbing mass are continuously recorded by the QMS. Examples of typical signals obtained from the QMS are shown in Figure \ref{Graph_raw} where the sudden increase and decrease of the mass signals are associated with the opening (irradiation) and closing (background) of the beamline shutter respectively. For each photon energy and each species, $I_X(E)$ is derived by computing the height of the mass signal with respect to the background level for the irradiation step considered. For molecules that could originate from the cracking of their desorbing parent molecules in the ionization chamber of the QMS, $I_X$ is corrected accordingly by using the cracking patterns available on the National Institute of Standards and Technology (NIST) chemistry Webbook. 
\\\\
The proportionality factor $k_X$ in equation \ref{Gamma_A} is first calibrated using the photodesorption of CO ($k_{CO}$): a VUV photodesorption spectrum of CO from pure CO ice is obtained by irradiating a pure CO ice (30 ML) previously grown at 15 K in the experiment and it is compared to a reference VUV photodesorption spectrum obtained at higher spectral resolution \citep{fayolle_co_2011}. $k_{CO}$ is deduced by matching the two spectra after taking into account the $\sim 1$ eV width of the photon energy profile used in the present study. The proportionality factors corresponding to neutral species other than CO are then computed by taking into account the differences in electron-impact ionization cross sections and apparatus functions of the mass filter, resulting in the following formula :
\begin{equation}
k_X = \frac{\sigma(CO^+ / CO)}{\sigma(X^+ / X)} \times \frac{AF(X)}{AF(CO)} \times  k_{CO} %\label{k_CO}
\end{equation}
where $AF(X)$ is the apparatus function of our QMS for the given species $X$ and $\sigma(X^+/X)$ is the non-dissociative electron-impact ionization cross section for the $X$ neutral species. These cross sections are taken at 70 eV and are available in the literature for CH$_3$CN \citep{zhou_total_2019}, C$_2$H$_6$ \citep{chatham_total_1984}, N$_2$ \citep{straub_absolute_1996}, HCN and CN \citep{pandya_electron_2012}, CH$_4$ \citep{Tian_1998} and CH$_3$ \citep{Tarnovsky_1996}. Due to the absence of available data, the non-dissociative electron-impact ionization cross section of radical CHCN, $\sigma(CHCN^+ / CHCN)$ was assumed to be equal to the one of CH$_3$CN, $\sigma(CH_3CN^+ / CH_3CN)$. The apparatus function $AF(X)$ describes the transmission of the quadripolar mass filter in the QMS for a fragment with a given mass $m$, and thus participates, together with the ionization cross section, to the detection efficiency of a given species. For our instrument, the apparatus function has been measured and follows a power law: $AF(X) = (m/28)^{(1.77)}$.
\\\\
The uncertainty associated with $I_X(E)$, $\delta I_X(E)$, depends on the signal-to-noise ratio and is estimated as the sum of the half-height of the signal noise in the background, and on the top of each peak for the corresponding irradiation step (see Figure \ref{Graph_raw}). The error bars associated with the photodesorption yields for each energy and each species, $\delta \Gamma_X(E)$, and displayed in Figures \ref{Graph_Pur}, \ref{Graph_mix} and \ref{Graph_ind} are derived from the following equation:
\begin{equation}
\delta \Gamma_X(E) = k_X \times \frac{\delta I_X(E)}{\phi(E)}
\label{dGamma_A}
\end{equation}
Equation \ref{dGamma_A} results in experimental error bars which depend on the photon energy. Consequently, these error bars are usually larger at high energy due to a lower photon flux. Finally, these error bars only reflect the quality of the recorded signals and do not take into account the uncertainty associated with the calibration of $k_X$, which is systemic. This uncertainty is not impacting the shape of the UV photodesorption spectra but only the uncertainty on the absolute values of the photodesorption yields. It mainly depends on the uncertainties associated with the apparatus function of the QMS and on the ionization cross sections $\sigma(X^+ / X)$. We estimate this uncertainty to be in the order of magnitude of 50$\%$ relative to the measured photodesorption yields.

\clearpage

\section{Attribution of the desorbing mass 28 channel using TPD} \label{AppB}

\begin{figure}[!ht]
\begin{center}
\resizebox{\hsize}{!}{\includegraphics{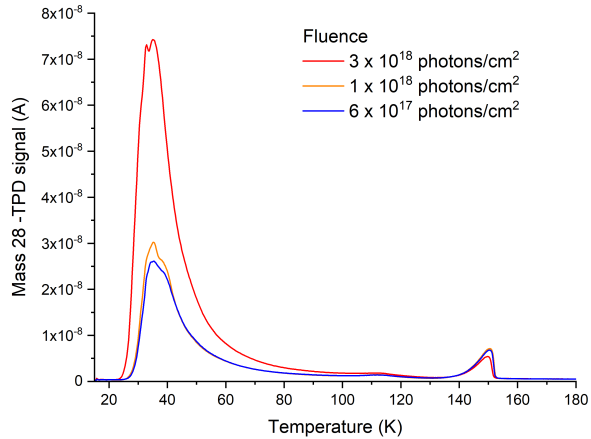}}
\caption{TPD signal on the mass 28 from pure CH$_3$CN ices as a function of the temperature after its irradiation by successive energy scans between 7 and 14 eV. The fluence received by the ice before TPD is also displayed. The TPD curves were recorded using a constant heating rate of 12 K/min. }
\label{Graph_tpd}
\end{center}
\end{figure}

As stated in the main text, each studied ices, after one or several irradiations, i.e. energy scans between 7 and 14 eV, were further probed using the Temperature Programmed Desorption (TPD) technique. To this end, the ices were warmed-up using a constant heating rate of 12 K/min, and the desorbed species were detected concomitantly using mass spectrometry. TPD has been in particular used in order to identify the origin of the mass channel 28 detected during photodesorption. Figure \ref{Graph_tpd} presents several TPD curves asociated with the mass channel 28 performed from irradiated CH$_3$CN ices. As one can see from these curves, the majority of the desorption occurs below 40 K, with an onset of desorption at around 25-27 K. Such a low temperature desorption is usually characteristic of the desorption of very volatile species, and match very well the thermal desorption of a small quantity of N$_2$ deposited onto metallic substrates or water ice substrates (see e.g. work from \citealt{collings_laboratory_2004} and \citealt{smith_desorption_2016}). The other feature observed at higher temperature is in fact associated to the desorption of N$_2$ trapped into the CH$_3$CN bulk, which is suddenly released when the ice is sublimated.
\\\\
Since the 28 signal desorption rises with the irradiation fluence, we attributed this TPD feature to the accumulation of N$_2$ onto or into the ice during the irradiation. And thus we believe that the photodesorbed mass 28 signal should also be due to some N$_2$ desorption subsequent to its formation. It should be noted however that TPD is not in itself sufficient to be categorical on this attribution, since, for instance, C$_2$H$_4$ could also contribute to this mass 28 signal, even if it should be obseved at slightly higher temperature. Finally, any possible uncontrolled pollution by another species, such as CO, has been ruled-out by (i) blank experiments, (ii) the correlation with the photon fluence, and (iii) the fact that no CO vibration could have been detected using in-situ infrared spectroscopy on the irradiated samples. The mass signal 14 was not monitored during the TPD experiments, which prevents us from definitively concluding on the origin of the mass 28 thanks to a possible fragmentation of N$_2$ into N. A mass signal 14 was however detected during photodesorption experiments but it was high enough so that the possible fragmentation of N$_2$ into N in the ionization chamber of our QMS has no significative impact on its amplitude, which also prevents us from definitively concluding. The photodesorption of this mass 14 is therefore due to either CH$_2$ radical or atomic N photodesorption.  \\\\

\clearpage

\section{UV photodesorption yields of masses 39, 30, 28, 16 and 15 from binary ices at 15 K} \label{AppC}

\begin{figure}[!ht]
\begin{center}
\resizebox{\hsize}{!}{\includegraphics{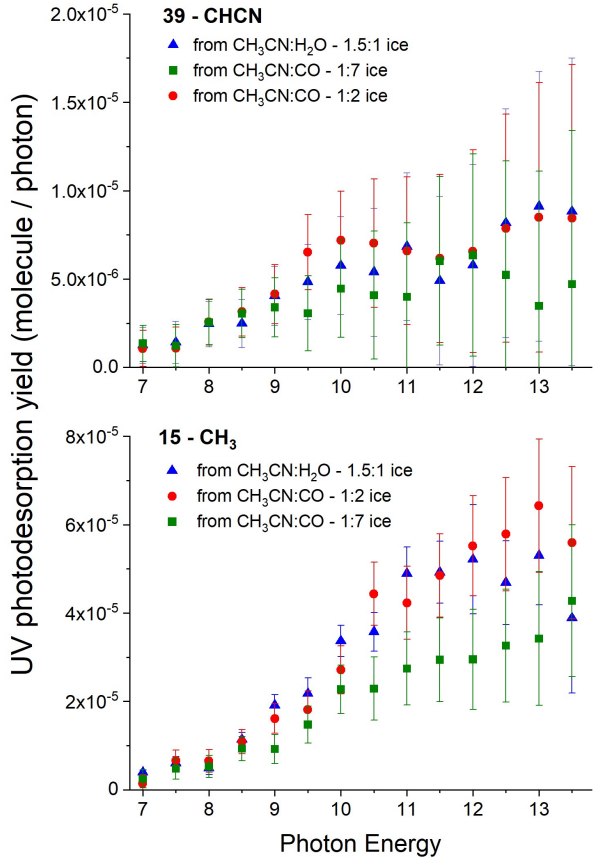}}
\caption{VUV photodesorption spectra, as a function of the incident photon energy, of CHCN$\bullet$ (mass 39) and CH$_3\bullet$ (mass 15) from mixed CH$_3$CN:CO and mixed CH$_3$CN:H$_2$O ices. }
\label{m_39_15}
\end{center}
\end{figure}

\begin{figure}[!ht]
\begin{center}
\resizebox{\hsize}{!}{\includegraphics{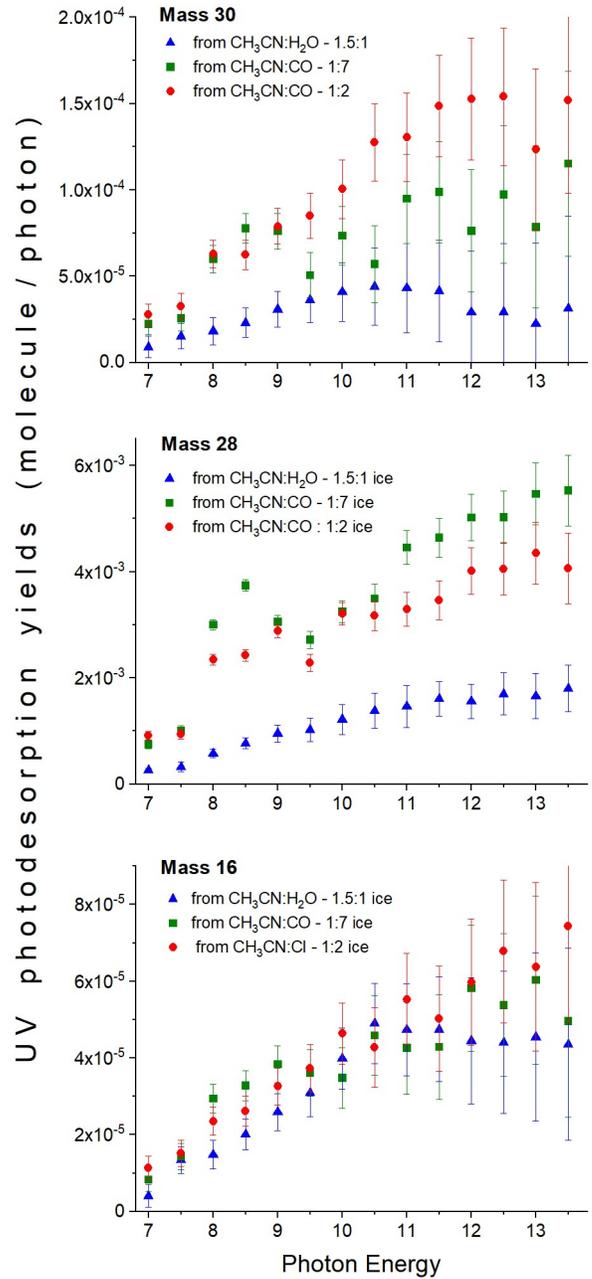}}
\caption{VUV photodesorption spectra, as a function of the incident photon energy, of masses 30, 28 and 16 from mixed CH$_3$CN:CO and mixed CH$_3$CN:H$_2$O ices. The attribution of species to these mass channels is not possible. This is discussed in Sec. \ref{sec:mix}.}
\label{m_302816}
\end{center}
\end{figure}

\end{document}